%% file: main.tex
\documentclass[11pt]{article}
\pdfoutput=1
\usepackage[mode=buildnew,subpreambles=true]{standalone}
\usepackage{fullpage}
\usepackage{amsmath}
\usepackage{amssymb}
\usepackage{amsthm}
\usepackage{authblk}

\usepackage{tikz}

\usepackage{hyperref}

\title{A Critique of Uribe's ``$\pe$ vs.\ $\np$''}

\author{Henry B. Welles\thanks{Supported in part by NSF grant CCF-2006496.}}
\affil{Department of Computer Science\\University of Rochester\\Rochester, NY 
14627, USA}

\date{May 2, 2022}

\newcommand{\naturalnumber}{\ensuremath{{\mathbb{N}}}}
\newcommand{\naturalnumberpositive}{\ensuremath{\mathbb{N}^+}}

\newcommand{\pe}{\mbox{\rm P}}
\newcommand{\np}{\mbox{\rm NP}}

\newcommand{\clique}{\mbox{\rm CLIQUE}}

\DeclareMathOperator{\bigOmega}{\Omega}

\newtheorem{theorem}{Theorem}

\begin{document}
	
\maketitle

\begin{abstract}
	In this critique, we examine the technical report by Daniel Uribe entitled 
	``P vs.\ NP'' \cite{uri:t:p-vs-np}. The paper claims to show an exponential 
	lower bound on the runtime of algorithms that decide $\clique$. We show 
	that the paper's proofs fail to generalize to all possible algorithms and 
	that, even on those algorithms to which the proofs do apply, the proofs' 
	arguments are flawed.
\end{abstract}

\section{Introduction} \label{sec:intro}

We give an overview and critique of Uribe's paper entitled ``P vs.\ NP'' 
\cite{uri:t:p-vs-np}. Uribe's paper claims to show an exponential lower bound 
on the runtime of every algorithm that decides $\clique$. To appreciate how 
important that advance would be, one must first recognize the significance of 
the $\pe$ vs.\ $\np$ problem 
and $\clique$'s relevance to it.

Many important problems fall into $\np$, the class of languages for which there 
exist ``fast'' membership-verification algorithms.
Some of these problems are ones that we would like to be able to solve quickly 
such as Job Sequencing, which is used heavily by airlines and carriers when 
planning trips. On the other hand, the security of Diffie-Hellman key exchange 
relies upon the difficulty of the discrete log problem 
\cite{dif-hel:j:diffie-hellman} (which has a language version in $\np$ 
\cite{ant-and-cec:b:open-in-math-cs}). As such, security researchers might 
prefer discrete log and other cryptographically significant $\np$ problems to 
remain infeasible.
The question of whether all $\np$ problems are decidable in polynomial time 
(i.e., whether all problems that have ``quickly'' verifiable solutions have 
``quick'' membership-testing algorithms) is referred to as the $\pe$ vs.\ $\np$ 
problem, and is considered the most important unsolved problem in computer 
science and, potentially, the entirety of applied mathematics. 
Of course, showing that a single $\np$ problem is not in $\pe$ is enough to 
show that $\pe \neq \np$. Since $\clique$ is an $\np$ problem 
\cite{kar:b:reducibility}, Uribe's paper showing that it cannot be decided in 
polynomial time (i.e., showing it is not in $\pe$) would be enough to prove 
that $\pe \neq \np$. We will show, however, that the paper's proofs are 
fundamentally flawed and that, as a result, they fail to establish the paper's 
central claim.

Section~\ref{sec:prelim} briefly summarizes the preliminaries required to 
discuss Uribe's paper. In Section~\ref{sec:summary}, we will first define the 
paper's core definitions and concepts, and then will provide a condensed 
version of the paper's core theorems\footnote{We note that the paper contains 
both ``claims'' and theorems. However, in this critique we will refer to both 
as theorems. Additionally, none of the theorems are numbered (or differentiated 
in any way) in Uribe's paper, so we will initially refer to them by the page on 
which they originally appear.} (and their proofs) that lead to its central 
claim. In Section~\ref{sec:critique}, we highlight the key errors in the 
paper's proofs and show how those errors result in the paper's failure to prove 
$\pe \neq \np$.

\section{Preliminaries} \label{sec:prelim}

We let $\naturalnumber = \{0,1,2,\ldots\}$ i.e., the set of natural numbers, 
and $\naturalnumberpositive = \{1,2,3,\ldots\}$ i.e., the set of positive 
natural numbers. Given $n,k \in \naturalnumber$ with $k \leq n$, the binomial 
coefficient on $\binom{n}{k}$ is defined as $\frac{n!}{k!(n-k)!}$.

\begin{figure}[h!]
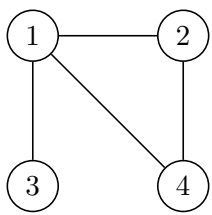

	\centering
	\includestandalone{simple}
	\caption{A visual representation of the graph $P$.}
	\label{fig:ex}
\end{figure}

An undirected graph is defined as a pair $(V,E)$ where $V$ is a finite set such 
that $V \subseteq \naturalnumberpositive$ and $E \subseteq \{\{a,b\} \mid a,b 
\in V\}$. Members of $V$ are termed vertices and members of $E$ are termed 
edges. For example, given a vertex set $V = \{1,2,3,4\}$ and an edge set $E = 
\{\{1,2\},\{1,3\},\{1,4\},\{2,4\}\}$, the resulting graph $P=(V,E)$ is pictured 
in Figure~\ref{fig:ex}.

Given a graph $G$, we use the notation $V(G)$ to refer to the set of vertices 
of $G$, $E(G)$ to refer to the set of edges of $G$, and let $n(G)$ denote the 
number of vertices in $G$. An edge $e$ and a vertex $v$ are said to be incident 
if $v \in e$. If two unequal vertices $a$ and $b$ are incident to the same edge 
$e$, then $a$ and $b$ are neighbors and are said to be connected; additionally, 
$e$ is said to be connecting $a$ and $b$. Given a graph $G$, a subgraph $H$ of 
$G$ is a graph such that $V(H) \subseteq V(G)$ and $E(H) \subseteq E(G)$. Note 
that as the set of possible edges of a graph is defined in terms of that graphs 
vertices, $H$ will contain a subset of the edges in $G$, restricted to those 
that are incident only to vertices in $V(H)$. A clique within a graph $G$ is a 
subgraph of $G$ in which every vertex is connected to every other vertex. When 
a clique contains $n$ vertices we will say it has size $n$ or we will simply 
refer to it as an $n$-clique. An example of a 3-clique is the subgraph 
containing the vertices 1, 2, and 4 (and all of the edges connecting them) in 
Figure~\ref{fig:ex}. The $\clique$ problem is the task of answering the 
following question: given a graph $G$ and $q \in \naturalnumberpositive$, does 
$G$ contain a $q$-clique?

A graph containing $n$ vertices can also be represented via an $n \times n$ 
matrix called an adjacency matrix. Within an adjacency matrix, the entry in the 
$i$-th row and $j$-th column has value 1 if and only if the graph corresponding 
to that matrix contains an edge connecting vertex $i$ and vertex $j$. To align 
with Uribe's paper, we assume there is an implicit connection between every 
vertex and itself. Note that this results in every entry on the main diagonal 
of an adjacency matrix having value 1. As an example, the adjacency matrix of 
the graph in Figure~\ref{fig:ex} is:
\[
\begin{bmatrix}
	1 & 1 & 1 & 1\\
	1 & 1 & 0 & 1\\
	1 & 0 & 1 & 0\\
	1 & 1 & 0 & 1\\
\end{bmatrix}
.
\]

\section{Understanding the Paper's Argument} \label{sec:summary}

Uribe's paper applies a decision-tree-based proof method to the $\clique$ 
problem. The paper uses this method to claim that a correct algorithm to decide 
$\clique$ must (in the worst case) traverse a number of interior nodes of the 
tree exponential in $n(G)$. Thus a correct algorithm for $\clique$ cannot have 
a polynomial runtime. We provide an overview of the key theorems and proofs 
underpinning the paper's method and then present the proof of the paper's 
central theorem.\footnote{We note that we do not always state the paper's 
theorems, definitions, and terms exactly as they originally appear, but instead 
use equivalent (but more common) language and notation.}

\subsection{Definitions and Concepts}

Uribe's paper centers around a method of using decision trees to compute lower 
bounds on the runtimes of algorithms. Given an algorithm, a decision tree for 
that algorithm is defined as a binary tree in which every interior node 
represents a binary comparison made by the algorithm and the connection of a 
node to its left or right child represents a decision made by the algorithm. 
Each leaf node of this decision tree represents an outcome of the algorithm. 
Uribe's paper uses sorting a sequence of integers as a demonstration of this 
approach; every interior node represents a comparison of two values in the 
sequence, and every leaf node represents a permutation to apply to the sequence 
in order to (ideally) sort it. 

The paper also uses the term solution set; this term is never formally defined. 
To account for this, we will now attempt to define it based off of the 
(minimal) context in which it is used in the paper. We define the solution set 
of an algorithm on a given input to be the set of all terminal states of the 
algorithm such that a solution for the input may have been found. As an 
example, the solution set of an algorithm to sort a given sequence of integers 
contains every permutation that could be applied to that input sequence (this 
is the case as for every permutation, there exists some sequence that it 
sorts). It is in this way that the leaf nodes of an algorithm's decision tree 
and members of its solution set are related; that is, when the algorithm 
reaches a leaf node, if it has found a solution for its input, then the 
algorithm's current state is in its solution set.

\subsection{The Argument}

The paper begins by providing a method of checking a graph for the presence of 
a given subgraph in polynomial time. This process is termed a subgraph 
comparison. The paper then presents the following theorem.

\begin{theorem}[\cite{uri:t:p-vs-np}, Theorem p.\ 22]
	\label{thm:min-comp}
	Let $G$ be a graph containing $n$ vertices and let $q \in 
	\naturalnumberpositive$ with $q \leq n$. The minimum number of subgraph 
	comparisons required for a deterministic algorithm to identify the presence 
	of a $q$-clique in $G$ is $\binom{n}{q}$.
\end{theorem}

\begin{proof}[Proof summary]
	As there are $n$ vertices in $G$, each potential $q$-clique is induced by 
	choosing $q$ of the vertices in $G$. Thus there are $\binom{n}{q}$ possible 
	$q$-cliques in $G$. As each $q$-clique represents a subgraph which may or 
	may not be present in $G$, there are $2^{\binom{n}{q}}$ possible states in 
	the solution set. Thus the minimum height of a decision tree spanning this 
	solution set is $\binom{n}{q}$.
\end{proof}

The paper then goes on to say that because we must perform a minimum of 
$\binom{n}{q}$ subgraph comparisons, and there are $\binom{n}{q}$ $q$-cliques 
in $G$, we can conclude that a linear search through the possible $q$-cliques 
of $G$ is an optimal algorithm.

Following this is a lengthy discussion to derive the well established result 
that, in the worst case (when $q=\frac{n}{2}$), $\binom{n}{q}$ is bounded below 
by an exponential function that is in  $\Theta(\frac{2^n}{n})$.\footnote{We 
note that the original paper only states that the lower bounding function is in 
$O(\frac{2^n}{n})$. However, as the lower bounding function used in the proof 
is $\frac{2^n}{n+1}$ which is clearly in $\Theta(\frac{2^n}{n})$, and as 
following proofs seem to rely on this tighter classification, we assume it is 
what the paper intended.}

We now present Theorems~\ref{thm:leaf} and~\ref{thm:main} which lead to a proof 
of the paper's central claim.

\begin{theorem}[\cite{uri:t:p-vs-np}, Theorem p.\ 35]
	\label{thm:leaf}
	Given a graph $G$ containing $n\geq 4$ vertices and a single $\lfloor 
	\frac{n}{2} \rfloor$-clique $Q$, the decision tree of every optimal 
	algorithm must have more than one leaf node identifying $Q$ as a subgraph 
	of $G$.
\end{theorem}

Although this theorem does not have a significant role in showing the
paper's claims, we will use it in Section~\ref{sec:critique} to highlight some 
of the key flaws within the paper's arguments.

\begin{theorem}[\cite{uri:t:p-vs-np}, Theorem p.\ 37]
	\label{thm:main}
	The runtime of every optimal algorithm that, given a graph $G$ and $q \in 
	\naturalnumberpositive$, returns the first instance of a $q$-clique is in 
	$\bigOmega(\frac{2^n}{n})$.
\end{theorem}

\begin{proof}[Proof summary]
	Consider the adjacency matrix of an $\frac{n}{2}$-clique $Q$. The part of 
	this matrix not relevant to the clique may be seen as another subgraph $M$ 
	containing $m=\frac{n}{2}$ vertices. By the same reasoning as in the proof 
	of Theorem~\ref{thm:min-comp}, the height of the decision tree of an 
	algorithm that will find $Q$ if it is present in a given graph is 
	$\binom{m}{\frac{m}{2}}$, and from the previous discussion on bounding 
	$\binom{n}{q}$, we know that this height is a function on $m$ in 
	$\bigOmega(\frac{2^m}{m})$. As $m$ is a scalar multiple of $n$, we have 
	that this function is also in $\bigOmega(\frac{2^n}{n})$.\footnote{We note 
	that as this summarizes the proof as it appears in Uribe's paper, it 
	inherits some of the flawed arguments of the original proof. These flaws 
	will be discussed in Section~\ref{sec:critique}.}
\end{proof}

The paper then states that, as a result of Theorem~\ref{thm:main} (and, it 
seems, implicitly Theorem~\ref{thm:min-comp}), $\clique \notin \pe$ and thus 
$\pe \neq \np$.

\section{Identifying the Error} \label{sec:critique}

This paper has a multitude of technical issues, some large with massive effects 
on the validity of the paper's claims, and some small with limited 
repercussions on the paper as a whole. For the sake of brevity we will not 
cover all of these issues, but will instead highlight the issues that have the 
largest and most direct impact on the paper's central claim.

\subsection{Definitions}

It is important to note how Uribe's paper defines the $\clique$ problem. It is 
done as follows: Given a graph $G$ and $q \in \naturalnumberpositive$, identify 
a $q$-clique in $G$ if one exists, and otherwise indicate that no $q$-clique 
exists in $G$. Furthermore, when proving the final claim that $\pe \neq \np$, 
the paper reads ``The clique problem, regardless of whether the search is for 
all cliques or just a single clique...'' \cite[p.\ 39]{uri:t:p-vs-np}. We note, 
however, that as we are discussing $\np$, $\clique$ must be considered to be a 
decision problem\footnote{A decision problem is a problem where there are two 
possible answers: either accept (i.e., there exists a $q$-clique in $G$) or 
reject (i.e., there does not exist a $q$-clique in $G$).} as it is defined by a 
language. So, for the rest of this paper, we will differ from Uribe's paper by 
treating $\clique$ as a decision problem. As such, we will critique the paper's 
arguments in terms of the definition of $\clique$ given in 
Section~\ref{sec:prelim}.

\subsection{Counting}

\begin{figure}[h!]
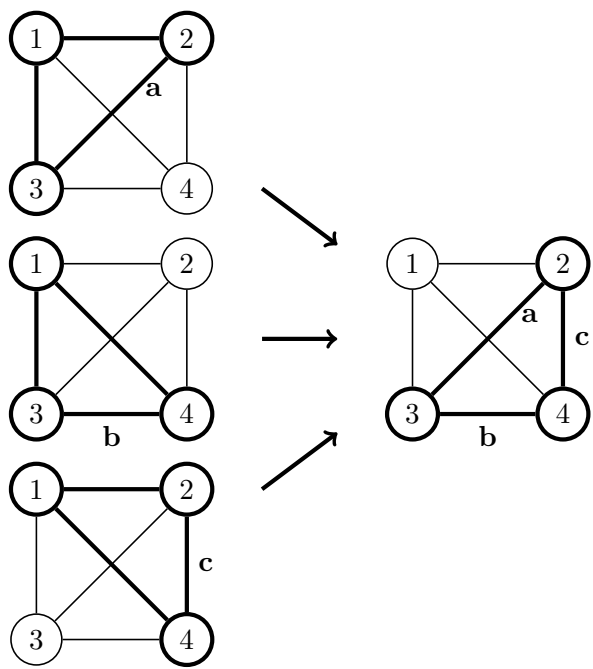

	\centering
	\includestandalone{depend}
	\caption{The presence of three cliques on the left (in bold) forces the 
	existence of edges a, b, and c. The existence of a, b, and c in turn forces 
	the presence of a fourth clique (shown on the right).}
	\label{fig:dep}
\end{figure}

In the proof of Theorem~\ref{thm:min-comp} and Theorem~\ref{thm:main} the paper 
asserts that if the number of possible cliques of size $q$ in a graph of size 
$n$ is $\binom{n}{q}$, then the solution set must contain at least 
$2^{\binom{n}{q}}$ possible outcomes. However, the reasoning used here is 
incorrect as the presence of one clique and the presence of another clique are 
not independent events. As seen in Figure~\ref{fig:dep}, it is entirely 
possible for the existence of certain cliques to imply the presence of another.

\subsection{Lower Bounds}

Even if this incorrect method of counting is used, and the results are simply 
taken as an upper bound, the following argument is still flawed. The paper, 
argues that if a solution set must contain at least $2^{\binom{n}{q}}$ 
outcomes, then the minimum height of a decision tree spanning the solution set 
is $\binom{n}{q}$. However, the paper incorrectly asserts that this implies a 
minimum number of $\binom{n}{q}$ subgraph comparisons must be performed. In 
contrast, what this is really saying is that, given a graph of size $n$, the 
maximum number of subgraph comparisons possibly needed to check for the 
presence of a clique of size $q$ is $\binom{n}{q}$. From this assertion the 
paper incorrectly concludes that a linear search though all cliques of size $q$ 
is optimal. This error seems to stem from the paper initially demonstrating the 
decision-tree-based technique in the context of analyzing sorting algorithms. 
However, there is an important distinction to be made between sorting 
algorithms and $\clique$. In sorting, one must consider every item in the 
sequence in order to ensure a correct sort. As a result one can assume that an 
algorithm must reach a leaf node in a  decision tree before terminating. So, 
the height of the decision tree is the worst case runtime of an algorithm, and, 
if this height is minimized, we have an algorithm with the best worst case 
runtime we can reasonably hope for.

However, $\clique$ is a very different story as a single subgraph comparison 
can result in the detection of a $q$-clique. Since this is a sufficient 
condition for an algorithm to accept (i.e., indicate that the graph does 
contain a $q$-clique) the execution of an algorithm can (and naturally should) 
stop there. This means that the height of a decision tree for an algorithm that 
decides $\clique$ does not represent the worst case runtime, but instead 
represents a trivially obtainable runtime that no efficient algorithm should 
ever exceed. There may, for example, be some clever method of initially 
reducing the number of comparisons one must make to an amount polynomial in the 
number of vertices in the input graph.

More importantly however, this paper makes unfounded assumptions about the 
types of algorithms we have at our disposal, focusing solely on decision-based 
algorithms. In the proof of Theorem~\ref{thm:leaf} for example, the paper 
states that when working with graphs, decisions may be based exclusively off of 
two things: (1) checking for connections between pairs of vertices and (2) 
performing subgraph comparisons. Not only does this ignore many of the other 
rich properties of graphs, but there are many other approaches to consider 
beyond those that work directly on the input graph. As an example, it may be 
the case that there exists a polynomial-time computable function that maps 
graphs to a field (e.g., the polynomials over the reals) and then easily 
computes a value indicating whether a clique is present or not.
Along the same lines, since Theorem~\ref{thm:main} only approaches the problem 
of searching for a $q$-clique from this narrow perspective, it ignores many 
other possibilities when establishing its bounds, for example the ability to 
reduce from the decision version to the search version of $\clique$.

\section{Conclusion} \label{sec:conclusion}

Due to vast oversimplifications and technical errors throughout, and despite 
drastically restricting the types of algorithms considered, Uribe's paper fails 
to prove any significant lower bound on the runtime of algorithms that decide 
$\clique$. Thus the paper fails to show $\pe \neq \np$.

\section{Acknowledgments} \label{sec:ack}

I thank
Michael C. Chavrimootoo,
Lane A. Hemaspaandra, and
Conor Taliancich
for their comments on earlier drafts of this paper. All remaining errors are
the responsibility of the author.

\bibliographystyle{alpha}
\bibliography{ref}
	
\end{document}

%% file: simple.tex
	\begin{tikzpicture}
		[
		auto, node distance =5cm, on grid,
		semithick,
		vert/.style ={circle, minimum size=2pt, draw, black, text=black},
		]
% 		\draw [help lines] (0,0) grid (2,2);
		\node [vert] (1) at (0,2) {1};
		\node [vert] (2) at (2,2) {2};
		\node [vert] (3) at (0,0) {3};
		\node [vert] (4) at (2,0) {4};
		\draw (1) -- (2);
		\draw (1) -- (3);
		\draw (1) -- (4);
		\draw (2) -- (4);
	\end{tikzpicture}

%% file: depend.tex
	\begin{tikzpicture}
		[
		auto, node distance =5cm, on grid,
		semithick,
		vert/.style ={circle, minimum size=2pt, draw, black, text=black},
		bold/.style ={ultra thick},
		]
% 		\draw [help lines]  (0,0) grid (8,8);
        \node [vert, bold]  (1c) at (0,8) {1};
		\node [vert, bold]  (2c) at (2,8) {2};
		\node [vert, bold]  (3c) at (0,6) {3};
		\node [vert]        (4c) at (2,6) {4};
		\draw       (1c) -- (4c);
		\draw       (2c) -- (4c);
		\draw       (3c) -- (4c);
		\draw[bold] (1c) -- (2c);
		\draw[bold] (1c) -- (3c);
		\draw[bold] (2c) -- (3c) node[very near start, below] {\textbf{a}};

		\node [vert, bold]  (1b) at (0,5) {1};
		\node [vert]        (2b) at (2,5) {2};
		\node [vert, bold]  (3b) at (0,3) {3};
		\node [vert, bold]  (4b) at (2,3) {4};
		\draw       (1b) -- (2b);
		\draw       (2b) -- (3b);
		\draw       (2b) -- (4b);
		\draw[bold] (1b) -- (3b);
		\draw[bold] (1b) -- (4b);
		\draw[bold] (3b) -- (4b) node[midway, below] {\textbf{b}};

		\node [vert, bold]  (1a) at (0,2) {1};
		\node [vert, bold]  (2a) at (2,2) {2};
		\node [vert]        (3a) at (0,0) {3};
		\node [vert, bold]  (4a) at (2,0) {4};
		\draw       (1a) -- (3a);
		\draw       (2a) -- (3a);
		\draw       (3a) -- (4a);
		\draw[bold] (1a) -- (2a);
		\draw[bold] (1a) -- (4a);
		\draw[bold] (2a) -- (4a) node[midway, right] {\textbf{c}};
		
		\draw[->, ultra thick] (3,6) -- (4,5.25);
		\draw[->, ultra thick] (3,4) -- (4,4);
		\draw[->, ultra thick] (3,2) -- (4,2.75);
		
		\node [vert]        (1a) at (5,5) {1};
		\node [vert, bold]  (2a) at (7,5) {2};
		\node [vert, bold]  (3a) at (5,3) {3};
		\node [vert, bold]  (4a) at (7,3) {4};
		\draw       (1a) -- (2a);
		\draw       (1a) -- (3a);
		\draw       (1a) -- (4a);
		\draw[bold] (2a) -- (3a) node[very near start, below] {\textbf{a}};
		\draw[bold] (3a) -- (4a) node[midway, below] {\textbf{b}};
		\draw[bold] (2a) -- (4a) node[midway, right] {\textbf{c}};
	\end{tikzpicture}